\documentclass[aps,prb,superscriptaddress,reprint]{revtex4-1} 
\usepackage{graphicx}
\usepackage{color}
\usepackage{braket}
\usepackage[separate-uncertainty=true]{siunitx}
\usepackage{graphicx}
\usepackage{bigints}
\usepackage{amsmath}
\usepackage{hyperref} 

\hypersetup{
    bookmarks=true,         
    unicode=false,          
    pdftoolbar=true,        
    pdfmenubar=true,        
    pdffitwindow=false,     
    pdfstartview={FitH},    
    pdftitle={title},    
    pdfauthor={Andreas Angerer},     
    pdfkeywords={keyword1} {key2} {key3}, 
    pdfnewwindow=true,      
    colorlinks=true,       
    linkcolor=red,          
    citecolor=blue,        
    filecolor=magenta,      
    urlcolor=cyan,           
}

\newcommand{\rom}[1]{\uppercase\expandafter{\romannumeral #1\relax}}

\begin{document}

\title{Superradiant Hybrid Quantum Devices} 

\author{Andreas Angerer}
\email{andreas.angerer@tuwien.ac.at}
\affiliation{Vienna Center for Quantum Science and Technology, Atominstitut, TU Wien, Stadionallee 2, 1020 Vienna, Austria}

\author{Kirill Streltsov}   
\affiliation{Vienna Center for Quantum Science and Technology, Atominstitut, TU Wien, Stadionallee 2, 1020 Vienna, Austria}
\affiliation{Present Address: Institute for Theoretical Physics, Universit\"at Ulm, 89069 Ulm, Germany}

\author{Thomas Astner}
\affiliation{Vienna Center for Quantum Science and Technology, Atominstitut, TU Wien, Stadionallee 2, 1020 Vienna, Austria}

\author{Stefan Putz}
\affiliation{Vienna Center for Quantum Science and Technology, Atominstitut, TU Wien, Stadionallee 2, 1020 Vienna, Austria}
\affiliation{Present Address: Department of Physics, Princeton University, Princeton, New Jersey 08544, USA}

\author{Hitoshi Sumiya}
\affiliation{Sumitomo Electric Industries Ltd., Itami, Hyougo, 664-0016, Japan}

\author{Shinobu Onoda}
\affiliation{National Institutes for Quantum and Radiological Science and Technology, 1233 Watanuki, Takasaki, Gunma 370-1292, Japan}

\author{Junichi Isoya}
\affiliation{Research Centre for Knowledge Communities, University of Tsukuba, 1-2 Kasuga, Tsukuba, Ibaraki 305-8550, Japan}

\author{William J. Munro} 
\affiliation{NTT Basic Research Laboratories, 3-1 Morinosato-Wakamiya, Atsugi, Kanagawa 243-0198, Japan}
\affiliation{National Institute of Informatics, 2-1-2 Hitotsubashi, Chiyoda-ku, Tokyo 101-8430, Japan}

\author{Kae Nemoto}   
\affiliation{National Institute of Informatics, 2-1-2 Hitotsubashi, Chiyoda-ku, Tokyo 101-8430, Japan}

\author{J\"org Schmiedmayer}
\affiliation{Vienna Center for Quantum Science and Technology, Atominstitut, TU Wien, Stadionallee 2, 1020 Vienna, Austria}

\author{Johannes Majer}
\affiliation{Vienna Center for Quantum Science and Technology, Atominstitut, TU Wien, Stadionallee 2, 1020 Vienna, Austria}
\affiliation{Wolfgang Pauli Institut, c/o Fak. Mathematik Univ. Wien}

\date{\today}

\begin{abstract}
Superradiance is the archetypical collective phenomenon where radiation is amplified by the coherence of emitters. It plays a prominent role in optics, where it enables the design of lasers with substantially reduced linewidths, quantum mechanics, and is even used  to explain cosmological observations like Hawking radiation from black holes. Hybridization of distinct quantum systems allows to engineer new quantum metamaterials pooling the advantages of the individual systems. Superconducting circuits coupled to spin ensembles are promising future building blocks of integrated quantum devices and superradiance will play a prominent role. As such it is important to study its fundamental properties in hybrid devices. Experiments in the strong coupling regime have shown oscillatory behaviour in these systems  but a clear signature of Dicke superradiance has been missing so far.  Here we explore superradiance in a hybrid system composed of a superconducting resonator in the fast cavity limit inductively coupled to an inhomogeneously broadened ensemble of nitrogen-vacancy (NV) centres.  We observe a superradiant pulse being emitted a trillion of times faster than the decay for an individual NV centre.  This is further confirmed by the non-linear scaling of the emitted radiation intensity with respect to the ensemble size. Our work provides the foundation for future quantum technologies including solid state superradiant masers.
\end{abstract}

\maketitle

Proposed by Dicke in 1954 \cite{dicke_coherence_1954}, superradiance is a collective effect enhancing the radiative decay dynamics of multiple excited emitters such that their decay is much faster than the individual emission rates - hence the name superradiance. Correlations which build up during the  decay lead to a non-linear scaling of the emitted radiation intensity with respect to the number of excited emitters \cite{gross_superradiance:_1982}. The effect of decoherence and dephasing in the system have to occur on a timescale longer than the relevant system dynamics such that coherence can be maintained throughout the decay,\cite{julsgaard_dynamical_2012} which has made superradiance difficult to observe. Experiments in a number of different systems have been carried out in a regime where dephasing effects are suppressed using a few qubits \cite{mlynek_observation_2014,eschner_light_2001,devoe_observation_1996,scheibner_superradiance_2007}, by taking advantage of special symmetries in the sample geometry \cite{gross_observation_1976,skribanowitz_observation_1973}, or in Bose-Einstein condensates \cite{inouye_superradiant_1999} and optical lattices \cite{ten_brinke_dicke_2015,norcia_superradiance_2016,meiser_prospects_2009}. A different approach is to use a resonator to decrease the mode volume of the electromagnetic field\cite{rose_coherent_2017,gross_maser_1979,kaluzny_observation_1983, putz_protecting_2014, mlynek_observation_2014} which enhances the coupling compared to dephasing mechanisms such as dipolar interaction induced broadening. To realize conditions close to those seen for superradiance in vacuum, the system needs to operate in the fast cavity limit, where the cavity decay rate is larger than all other time constants in the system, \cite{temnov_superradiance_2005,delanty_superradiance_2011,jodoin_superradiance_1974} or in the dispersive regime \cite{bennett_phonon-induced_2013,lambert_superradiance_2016}.

A suitable model system is an ensemble of negatively charged NV centres in diamond, coupled to a microwave resonator operating in the fast cavity limit.  Our measurements described below show both the non-linear emission of radiation from the fast cavity, while optical readout of the spin system confirms an enhanced decay rate a trillion times faster compared to the lifetime of a single NV centre. These observations are clear signatures of superradiance.

\begin{figure}[ht!]
\includegraphics[width=0.93\columnwidth]{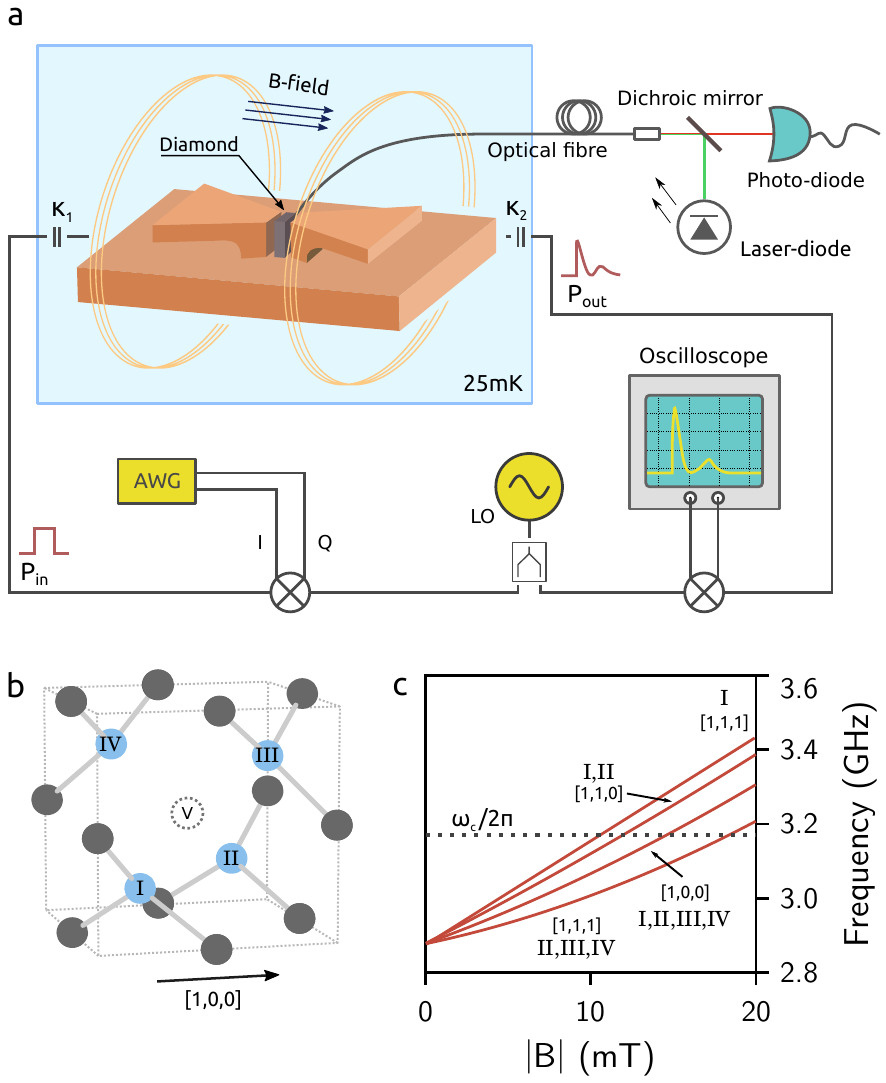}
\caption{\textbf{Experimental setup. a,} Our hybrid system consists of a 3D lumped element resonator (rendering without the top lid and the side walls) bonded with a diamond sample in combination with the laser readout scheme and the microwave setup operating at \SI{25}{\milli\kelvin} in a dilution refrigerator. A 3D Helmholtz coil configuration provides magnetic fields in arbitrary directions with field strengths up to \SI{200}{mT}. A fibre is glued to one side of the diamond sample to apply optical readout pulses and collect the scattered fluorescence, with a dichroic mirror to separate them. The microwave setup consists of a standard autodyne detection scheme with two mixers for (de-)modulation of the I/Q-signal. {\bf b,} The diamond lattice shows four different orientations for a given vacancy, resulting in four possible different NV directions (labelled \rom{1}-\rom{4}). {\bf c,} The Zeeman tuning of the NV levels by an external magnetic field is given by the projection of the magnetic field on the spins principal axis. This allows to bring either 1, 2, 3 or 4 NV subenssembles into resonance with the cavity mode ($\omega_c$), by applying appropriate magnetic fields in the [1,1,1], [1,1,0] and [1,0,0] direction.}
\label{fig:fig1}
\end{figure}

In our experiment, as illustrated in Fig.~\ref{fig:fig1}, we use a dense ensemble of NV centres with a narrow spectral linewidth of  $\gamma_\perp^*/2\pi\approx\SI{2.7}{MHz}$ (FWHM) in order to reduce dephasing effects originating from inhomogeneous broadening. The sample is placed in a 3D lumped element resonator\cite{angerer_collective_2016} with a fundamental frequency of $\SI{3.18}{GHz}$ and a cavity linewidth of $\kappa/2\pi=\SI{13.8}{MHz}$ (FWHM) corresponding to a quality factor of $Q=230$. The resonator focuses the magnetic field such that the spins are homogeneously coupled to the cavity mode with almost no spatial dependence on the coupling rate. This allows us to perform coherent operations on the entire spin ensemble ($\sim10^{16}$ spins) and achieve inversion using short microwave pulses, as illustrated in Fig.~\ref{fig:fig1}. Operating in the fast cavity limit ensures that the cavity increases the effective coupling to the spin sytem, but at the same time realizes conditions similar to superradiance in vacuum \cite{temnov_superradiance_2005}.  To meet the requirements necessary for the fast cavity limit, we ensure that the photon lifetime in the cavity of $\tau\approx\SI{11}{ns}$ is shorter than any of the occuring system dynamics.

\begin{figure}[ht!]
\includegraphics[width=0.93\columnwidth]{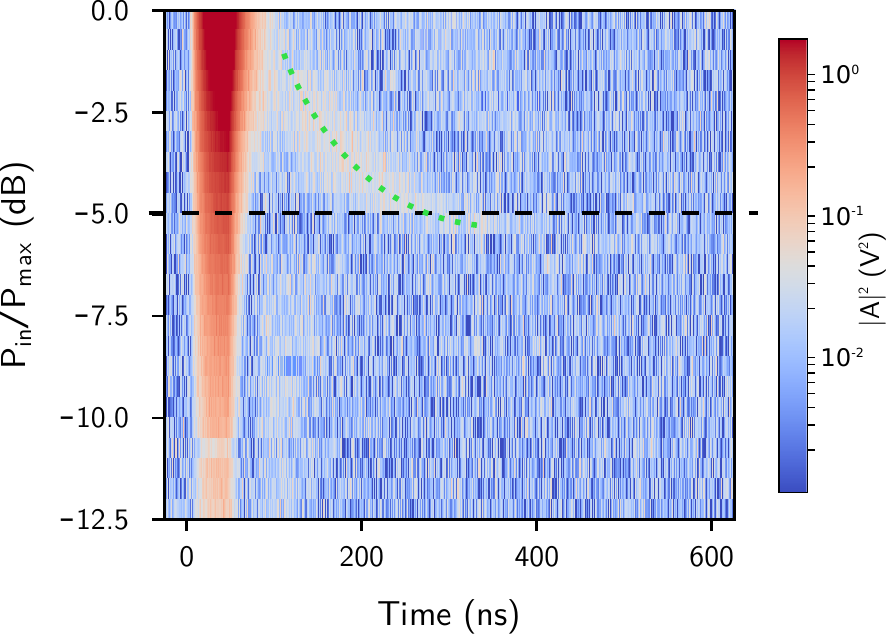}
\caption{{\textbf {Cavity response under varying drive powers.}} The response after applying a coherent microwave pulse of \SI{50}{ns} with three NV sub-ensembles in resonance with the cavity. The colour bar shows the emitted intensity $|A|^2$. At a power level ($\approx\SI{-5}{dB}$ attenuation of the maximum power, indicated with a black dashed line) we achieve maximum inversion and observe a pulse emitted from the cavity (green dotted line). For larger powers the delay of this emitted pulse becomes shorter since excess photons in the cavity lead to stimulated emission.}
\label{fig:fig2}
\end{figure}%

\begin{figure}[h!]
\includegraphics[width=\columnwidth]{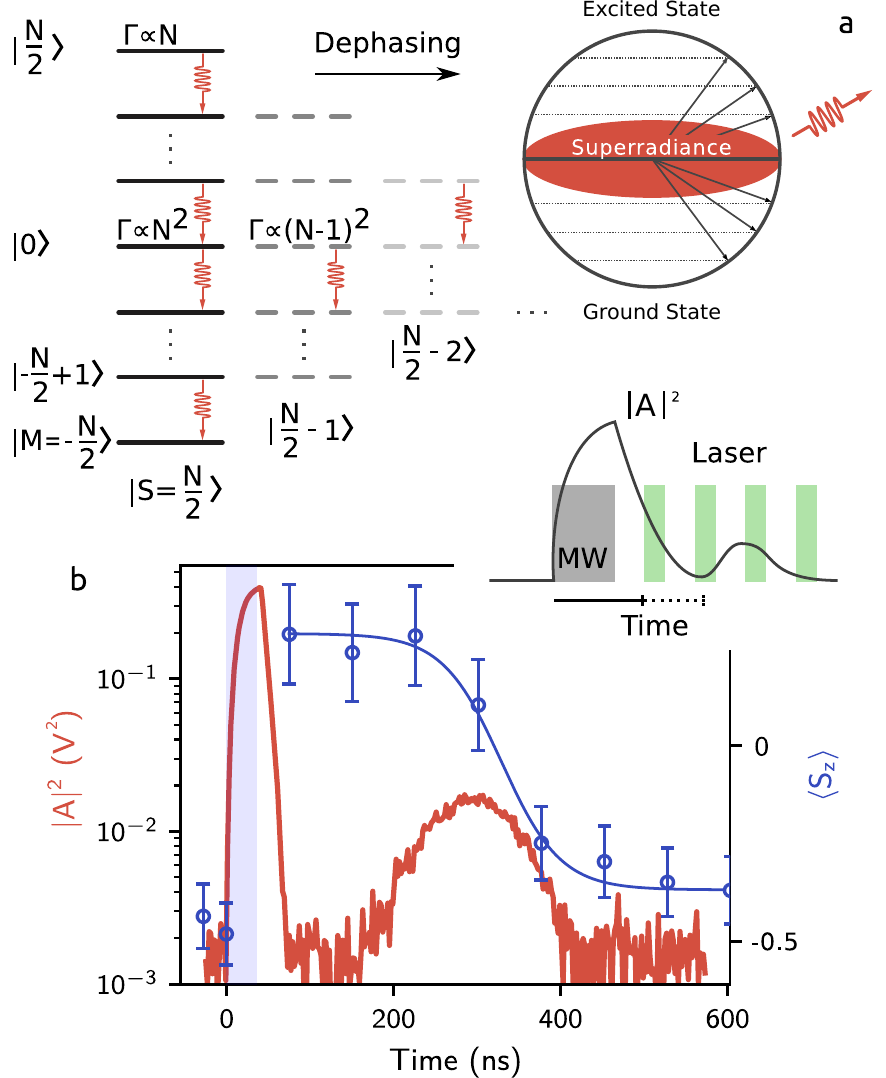}
\caption{\textbf {Dynamics of the superradiant decay. a,} Close to the region where the inversion $\braket {S_z}$ is zero and the spin quantum number is maximum, correlations in the spin system lead to an enhanced photon emission rate $\Gamma\propto N^2$ (note that we assume a $S=1/2$ system here). Uncorrelated emission $\Gamma\propto N$ governs the decay for the excited spin system. Dephasing decreases the spin quantum number and evolves the system out of the purely symmetric subspace. However, superradiance also occurs in these partly dephased inner shells, but the fully symmetric ground state is not reached anymore after the decay. Further, the number of photons emitted during the superradiant decay becomes smaller. \textbf {b,}  The red measurement curve shows a detailed view on the trace for the emitted photon intensity with three NV subensembles in resonance with the cavity and where the inversion of the spin ensemble is maximum (black dashed line in Fig.~\ref{fig:fig2}). The shaded area is the time for which the excitation drive is turned on. Shown in blue is the dynamics of the spin inversion, measured using the optical transition of the NV centre and the inversion polarization normalized to the number of spins. The inset shows the measurement sequence with MW excitation and optical readout pulses. After maximum inversion is reached the spins remain in a metastable state until fluctuations lead to a stimulated superradiant decay. This is accompanied by a burst of photons that builds up in the cavity mode. The blue solid line represents a fit of the fluorescence data according to a hyperbolic tangent\cite{gross_superradiance:_1982}.}
\label{fig:fig3}
\end{figure}
\begin{figure}[ht!]
\includegraphics[width=\columnwidth]{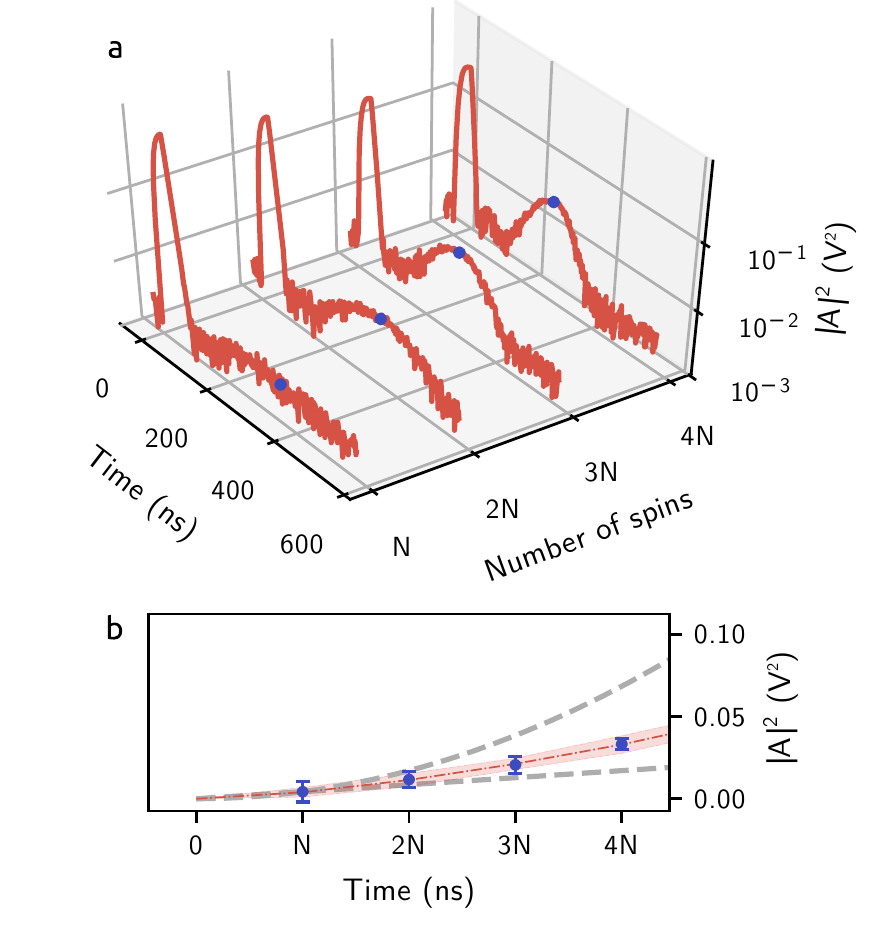}
\caption{ \textbf {Non-linear scaling of the emitted radiation intensity. a,} By bringing either $N$, $2N$, $3N$ or $4N$ spins into resonance with the cavity mode we are able to measure the dependence of the emitted photon intensity with respect to the number of spins. Depicted here are the traces for each of these cases where maximum inversion is reached, and a microwave drive duration of \SI{50}{ns}. \textbf {b,} By measuring the maximum value of the emitted intensity we observe  a non-linear scaling as $| A|^2\propto N^{1.52}$. The dashed grey lines show $N$ and $N^2$ scaling, respectively.}
\label{fig:fig4}
\end{figure}

We begin our exploration of this hybrid system by driving the spins with a $\SI{50}{ns}$ long pulse and measure the emitted radiation intensity from the cavity as a function of the input power as presented in Fig.~\ref{fig:fig2}. For small power levels the system primarily remains in the ground state, corresponding to the behaviour in the low-excitation (boson-like) regime. For a certain threshold power we achieve maximum inversion, indicated by the black dashed line in Fig.~\ref{fig:fig2}. At this point the system is in a metastable state - comparable to a classical pendulum turned upside down. The coupling of the spin system to the cavity mode is suppressed and it remains in this state for an extended period of time. In this metastable state the excitations reside in the spin system with the cavity mode empty. Approximately $\SI{300}{ns}$  after we switch off the microwave excitation, we observe a burst of photons exiting the cavity, depicted as red measurement data in Fig.~\ref{fig:fig3}. If we further increase the input power, the delay of this photon burst becomes shorter, because excess photons from the drive pulse keep interacting with the spin ensemble. This stimulates the emission of photons and drives the spins earlier out of their metastable state. 

Since superradiance is the enhanced coherent decay of the inverted spin system, it is instructive to measure the polarization inversion of the ensemble during this decay. The NV centre posesses an optical transition which enables a direct measurement of the spin polarization by optically detected magnetic resonance (ODMR)\cite{doherty_nitrogen-vacancy_2013,gruber_scanning_1997}. 
We implement ODMR in our experiment at ultra-low (mK) temperatures by illuminating parts of the sample using a \SI{20}{ns} long optical pulses delivered through an optical (multimode) fibre and collect the scattered fluorescence with the same fibre (see Methods for details). The fluorescence level then gives a direct measurement for $\braket {S_z}$. By varying the time-delay of the laser readout pulse with respect to the microwave pulse set for maximum inversion of the spin system, a time resolved measurement of the inversion is obtained. We measure the microwave cavity output and the scattered fluorescence simultaneously (details of the measurement are presented in the Methods section). As can be seen from the blue measurement data in Fig.~\ref{fig:fig3}, the inverted spin ensemble stays in its metastable state for approximately $\SI{200}{ns}$ before thermal and vacuum fluctuations stimulate the decay of the spins and the emission of microwave photons in a characteristic superradiant burst, as depicted in red in Fig.~\ref{fig:fig3}b. During drive and decay dephasing evolves some spins into a subradiant state (see Fig.~\ref{fig:fig3}a). This has the effect that we do not reach full inversion after the drive pulse and the spins do not decay back into the fully symmetric ground state after the superradiant decay with five percent incoherent inversion remaining. Dephasing also reduces the number of photons emitted during the superradiant burst. These dephasing effects are evident from the fluorescence level that does not reach the initial level after the superradiant burst (Fig.~\ref{fig:fig3}b). The remaining incoherent excitation  decays either through single spin Purcell-enhanced sponteaneous emission \cite{bienfait_controlling_2016} ($\Gamma_\mathrm P/2\pi\approx\SI{8e-10}{\hertz}$) or by longitudinal relaxation mediated by spin-phonon interactions \cite{astner_solid-state_2017} ($\gamma_\parallel/2\pi\approx\SI{3e-5}{\hertz}$), with the latter one dominating in our solid state spin system. 

We can study the non-linear scaling of the emitted photon radiation intensity, as expected from Dicke superradiance, by tuning either 1, 2, 3 or 4 NV subensembles in resonance with the cavity mode and thus changing the number of spins coupled from $\sim\SIrange{0.38}{1.5e16}{}$. 
In Fig.~\ref{fig:fig4}a we show the corresponding measurement data for the emitted intensity where the delay of the superradiant burst is maximum for each of these four cases.  By determining the maximum values of the intensities of the superradiant bursts and plotting them as a function of the number of spins coupled (Fig.~\ref{fig:fig4}b) we observe a dependence as $I\propto N^{1.52}$. This non-linear scaling as expected from a superradiantely enhanced decay\cite{gross_superradiance:_1982} clearly demonstrates the superradiant nature of the emitted photon burst.

Our system can be described by the driven Tavis-Cummings Hamiltonian \cite{tavis_exact_1968} which gives the dynamics of $N$ spins coupled to a quantized field mode.
In the fast cavity limit it is straightforward (details in the Method section) to derive an equation of motion for the inversion of the spins as, 

\begin{align}\label{eq:adiabatic}
\dot{\braket{S_z}} = -\frac{4g \eta}{\kappa}\braket{S_y}-\gamma_\parallel\left(\braket{S_z}+N\right)-\frac{4g^2}{\kappa}\braket{S_+S_-},
\end{align}
where $\eta$ is the drive amplitude, $\kappa$ the decay rate of the cavity intensity, $g$ the coupling rate of a single spin to the cavity mode, $\gamma_\parallel$ the longitudinal decay rate and the collective spin operators $S_{x, y, z}$, $S_{\pm}$.

Analyzing Eq.~(\ref{eq:adiabatic}) it is apparent that the cavity provides an enhancement of the sponteneous emission for the undriven system ($\eta=0$), known as the Purcell factor \cite{e._m._purcell_proceedings_1946}. This Purcell factor $4g^2/\kappa$ is increased by emerging correlations in the spin system during the decay of the inverted spin ensemble with a maximum proportionality of  $\braket{S_+S_-}\propto N^2$ (derived in the Method section). For our typical system parameters this enhances the small Purcell factor of $\Gamma_\mathrm P/2\pi\approx\SI{8e-10}{\hertz}$ by $\sim16$ orders of magnitude to a value of several \si{MHz}. The enhancement is clearly visible from the measurement data as displayed in blue in Fig.~\ref{fig:fig3}b, where we observe a spin decay with a maximum rate of $\Gamma_\mathrm{sr}/2\pi\approx\SI{17}{MHz}$, much faster than the Purcell enhanced decay of a single spin.

These emerging correlations also account for the fact that the intensity of the superradiant radiation emitted by the decaying spin system $I \propto \frac{g^2}{\kappa}\braket{ S_+S_-}$  scales non-linearly as the number of spins coupled to the cavity mode changes \cite{temnov_superradiance_2005,jodoin_superradiance_1974}. The non-linear scaling of the emitted radiation intensity, as illustrated in Fig.~\ref{fig:fig4}b, together with the enhanced decay of the spins as derived in Eq.~(\ref{eq:adiabatic}) and shown in Fig.~\ref{fig:fig3}, are both indicators for the superradiant nature of the emitted light.

Pure $N^2$ scaling is only expected for the limit of emitters without a cavity mode and no spectral broadening. However, as we are working in a cavity the scaling exponent becomes smaller than two. Further, this is enhanced by experimental imperfections, such as slightly misaligned magnetic fields for more than one subensemble in resonance with the cavity mode, which leads to more dephasing for more than one subensemble coupled. By ensuring that the cavity linewidth surpasses all other time constants and minimizing experimental imperfections, we still observe a non-linear scaling as expected from Dicke superradiance. The fact that the fast cavity limit is only a first order approximation also accounts for the occurence of a faint second pulse in Fig.~\ref{fig:fig2} for large drive powers. Photons that are emitted during the first superradiant pulse don't exit the cavity immediately and instead get partially reabsorbed by the spin ensemble again creating a superradiant burst.

Our studies described above clearly demonstrate Dicke superradiance in a hybrid quantum system embedding a macroscopic solid state spin ensemble, by measuring both the non-linear scaling of the emitted radiation and the enhanced decay of the spins. Especially the implementation of optically detected magnetic resonance techniques shows that protocols that rely on the optical transition in the NV centre are feasible even at low temperatures \cite{grezes_storage_2015}. This opens up new perspectives for solid state hybrid quantum systems employing both the full coherent microwave and optical control,  and shows the versatility of these types of systems. Our experiments show that the NV centre is a suitable candidate for quantum technologies that rely on a superradiant enhancement, such as the NV based superradiant maser by pumping of the optical transition instead of driving the microwave transition directly \cite{bohnet_steady-state_2012,jin_proposal_2015}. The superradiantly enhanced decay of the spins back into the ground state increases the repetition rate of the experiment by many orders of magnitude, which allows to perform high-sensitivity measurements \cite{weiner_superradiant_2012,acosta_diamonds_2009,bienfait_reaching_2016} without relying on the small spontaneous emission rate at low temperatures to polarize the spins back into the ground state. 

\textbf{~\\Acknowledgements}\\
We would like to thank Dmitry Krimer, Matthias Zens, Stefan Rotter and Helmut Ritsch for helpful discussions and Georg Wachter for help with the setup of the laser system. The experimental effort has been supported by the Top-/Anschubfinanzierung grant of the TU Wien and the JTF project: The Nature of Quantum Networks (ID 60478). A.A. and T.A. acknowledge support by the Austrian Science Fund (FWF) in the framework of the Doctoral School “Building Solids for Function” Project W1243. K.N. acknowledges support from the MEXT KAKENHI Grant-in-Aid for Scientific Research on Innovative Areas “Science of hybrid quantum systems” Grant No. 15H05870. J.I. acknowledges support by the Japan Society for the Promotion of Science KAKENHI Grant No.26220903 and Grant No.17H02751.

\newpage
\textbf{~\\Methods:}\\\\
{\textbf{Spin system}}\\
The negatively charged nitrogen-vacancy centre is a paramagnetic point defect centre in diamond, consisting of a nitrogen atom replacing a carbon atom and an adjacent lattice vacancy. Two unpaired electrons form a spin $S=1$ system, which can be described by a simplified Hamiltonian of the form $H=\hbar DS_z^2+\hbar\mu B_z S_z$ with a zero field splitting parameter $D/2\pi=\SI{2.878}{GHz}$ and a gyromagnetic ratio of $\mu/2\pi=\SI{28}{MHz\per mT}$. The splitting corresponds to a temperature of $D/\hbar k_b=\SI{138}{mK}$ which allows to thermally polarize the spins to the ground state at the refrigerator base temperature of $\approx\SI{25}{mK}$ with more than $99\%$ fidelity. The diamond lattice possesses four different crystallographic orientations (shown in Fig.~\ref{fig:fig1}b), which results in four different possible directions for the NV centre in the $\langle1,1,1\rangle$ directions. Each of these four vectors shows the same abundance of NV centres all exhibiting the same properties. By applying magnetic fields in the $[1,0,0]$ direction the external magnetic field projection onto the NV axis is equal for all four sub-ensembles, which allows us to tune all of them into resonance with the cavity mode. By applying appropriate magnetic fields in the $[1,1,1]$ direction, we can bring either only one or three sub-ensembles into resonance with the cavity depending on the magnetic field strength. Two sub-ensembles can be brought into resonance by applying magnetic fields in the $[1,1,0]$ direction.
\\\\
{\textbf{NV Sample}}\\
The sample is a type-Ib high-pressure, high-temperature (HPHT) diamond crystal with
an initial nitrogen concentration of \SI{50}{ppm}. In order to create lattice vacancies, the sample is irradiated with electrons of an energy of $\SI{2}{MeV}$ at $\SI{800}{\celsius}$ and is subsequentely annealed multiple times at $\SI{1000}{\celsius}$. The total electron dose was $\SI{5.6e18}{\centi\meter^{-2}}$. This gives a total NV density of \SI{13}{ppm} and an inhomogeneously broadened linewidth of $\gamma_\mathrm{inh}/2\pi\approx\SI{2.7}{\mega\hertz}$ (FWHM).. This value is fundamentally limited by the hyperfine coupling to the nuclear spin of nitrogen (\SI{2.3}{MHz}). The small spectral broadening allows us to increase our signal to noise ratio and number of photons emitted during the superradiant decay. Further, excess nitrogen P1 centres ($S=1/2$), uncharged $\mathrm{NV}^0$ and naturally abundand $^{13}\mathrm C$ nuclear spins serve as a source of decoherence and inhomogeneous broadening. The electron irradiation was performed using a Cockcroft-Walton accelerator in QST, Takasaki (Japan). The linewidth was measured using a dispersive measurement scheme relying on the dispersive shift of the microwave resonator coupled to a polarized spin ensemble \cite{amsuss_cavity_2011}.
\\\\
{\textbf{Spin Decay}}\\
The longitudinal decay of the NV ensemble $\gamma_\parallel$ is quantified using a dispersive measurement scheme, and exhibits the small value of \SI{3e-5}{\hertz}, mediated by spin-phonon interactions\cite{astner_solid-state_2017}. This value is many orders of magnitude larger than the Purcell factor, which on resonance is given by
\begin{align}
\Gamma_P=4\frac{g^2}{\kappa}.
\end{align}
With an estimated single spin coupling strength of $g/2\pi=\SI{72}{\milli\hertz}$ and a cavity linewidth of $\kappa /2\pi=\SI{13.8}{\mega\hertz}$, this computes to a value of $\Gamma_\mathrm P/2\pi=\SI{7.5e-10}{\hertz}$. 
\\\\
{\textbf{Hybrid System}}\\
We use a 3D lumped element resonator machined out of oxygen free copper that allows to couple homogeneously to the whole diamond sample with a root mean square deviation of the coupling rate of $\approx1.5\%$ \cite{angerer_collective_2016}. Our hybrid system is formed by bonding the NV sample within the 3D lumped element resonator. The collective coupling strength to the entire spin ensemble changes from $\SIrange{3.1}{6.2}{MHz}$ depending on the number of subensembles coupled to the cavity mode. We operate the resonator in the largely overcoupled regime $\kappa_1\gg\kappa_2\gg\kappa_{int}$ with the input coupling much larger than the output coupling. This increases the number of photons in the cavity and consequentely reduces the time it takes to perform a $\pi$-pulse. Further, it allows us to perform the experiments in the fast cavity regime, such that the cavity linewidth surpasses all other time constants in the system ($\kappa>\Theta_j, \gamma_\perp, \gamma_\parallel,\sqrt{N}g$). 
\\\\
{\textbf{Photon Intensity Measurements}}\\
We measure the emitted photon intensity by performing transmission measurements using a autodyne detection scheme. The carrier frequency is split into two paths with one serving as the signal going into the cryostat. It is modulated using a fast arbitrary waveform generator (Tabor WW2182B) with \SI{2}{GS s^{-1}} sampling frequency to produce short microwave pulses. The other path is used as a reference signal used to demodulate the signal exiting the crysotat. The demodulated signal,  with the intensity proportional to the emitted photon flux, is then recorded by a fast digitizer card (Acqiris U1082A) with \SI{5}{GS s^{-1}} sampling frequency.
\\\\
{\textbf{Optical Fluorescence Measurement}}\\
To get a relative measure for the inversion  of the spin system, we make use of the optical transition of the NV centre that gives a direct way to measure the expectation value for the $\braket{S_z}$ component. We glue a multimode fibre to one side of the diamond sample and illuminate it with short (i.e. \SI{20}{ns}) green \SI{526}{nm} laser pulses using a high bandwidth laser-diode (LD-520). The scattered red light is collected using the same fibre, and filtered using a dichroic mirror reflecting the ingoing green light and transmitting the scattered red light. An optical longpass filter further filters the scattered red light which is then measured by an avalanche photo-diode (Thorlabs ADP110A). Because the coupling rate to the cavity mode shows almost no spatial dependence we can assume that optical readout of a small part of the sample is equivalent to the readout of the whole ensemble. We record the fluorescence for several hundred repetitions for different time delays of the readout pulse with respect to the microwave excitation pulse. For a spin system that is fully polarized in the ground state this gives the highest fluorscence, while the maximally inverted spin system exhibits the lowest fluorescence. By comparing the fluorescence values of this scattered light we get a relative measure for the $\braket{S_z}$ component. 
\\\\
{\textbf{Modelling: Equations of Motion and the Adiabatic Elimination in the Fast Cavity Limit}}\\
In order to derive the equation of motions in the fast cavity limit we begin with a driven Tavis-Cummings Hamiltonian for $N$ spins 
in the rotating frame
\begin{align}\label{eq:tavis}
\begin{split}
\mathcal H= \hbar \Delta_c a^\dag a + \frac{\hbar}{2}\sum_{j=1}^N \Delta_s^j\sigma_z^j + \\
+ i \hbar g \sum_{j=1}^N\left(a^\dag\sigma_-^j-\sigma_+^j a\right)&+i \hbar \eta\left(a^\dag-a\right),
\end{split}
\end{align}

with $\sigma^j_z$, $\sigma^j_\pm$ the Pauli-$z$ and raising/lowering operator for the $j^\mathrm{th}$ spin and $a,a^\dag$ the bosonic creation and annihiliation operators. Note that the collective spin operators in the main text are defined as  $S_{x, y, z}  = \frac{1}{2}\sum_{j=1}^N \sigma^j_{x,y,z}$ and  $S_{\pm} = \sum_{j=1}^N \sigma^j_{\pm}$. Further, $\Delta_s^j$ is the detuning of the $j^\mathrm{th}$ spin with respect to the cavity resonance frequency accounting for the inhomogeneous broadening of the spectral spin linewidth.
In the fast cavity limit we adiabatically eliminate the cavity mode, and arrive at a Hamiltonian which only contains spin operators, as

\begin{align}\label{eq:ham}\mathcal H_{at}&= \frac{\hbar}{2} \sum_{j=1}^N\Delta_s^j \sigma^j_z - 2\hbar \frac{g\eta}{\kappa}\sum_{j=1}^N(\sigma^j_++\sigma^j_-),
\end{align}
Using the Lindblad master equation, introducing loss channels for the cavity mode and the spins as

\begin{align}\label{eq:lindblad}
\begin{split}
\dot\rho= &-\frac{i}{\hbar}\left[\mathcal H, \rho\right]+ \\
+&\kappa\left(a\rho a^\dag - \frac{1}{2}\left(a^\dag a\rho - \rho a^\dag a\right)\right) \\
+&\gamma_\parallel\sum_{j}^N\left(2\sigma_-^j\rho\sigma_+^j - \sigma_+^j\sigma_-^j\rho-\rho\sigma_+^j\sigma_-^j\right) \\
+&\gamma_\perp\sum_{j}^N\left(\sigma_z^j\rho\sigma^j_z-\rho\right)
\end{split}
\end{align}
 we derive an equation of motion for the cavity mode operator as
\begin{align}
\dot a = -i\Delta_c a+ g\sum_{j=1}^N \sigma_-^j+\eta-\frac{\kappa}{2} a.
\end{align}
In the fast cavity limit we assume $\dot a = 0$, since the presence of the coupled spin system in this limit does not alter the cavity amplitude which remains unchanged. The resulting expression for the mode operator, assuming the cavity mode in resonance with the probe frequency ($\Delta_c=0$), can be resubstituted into Eq.~(\ref{eq:tavis}) which then leads to a Hamiltonian only for the atomic operators, as given in Eq.~(\ref{eq:ham}). The losses in the cavity in the Lindblad superoperator from Eq.~(\ref{eq:lindblad}) can then be expressed by the atomic operators, as

\begin{align}
\mathcal L_\mathrm{cav}=&2\frac{g^2}{\kappa}\sum_{j,k}^N\left(2\sigma_-^j\rho\sigma_+^k - \sigma_+^j\sigma_-^k\rho-\rho\sigma_+^j\sigma_-^k\right).
\end{align}
From this we can derive the equation of motions for the spin inversion as given in Eq.~(\ref{eq:adiabatic}).
\\\\
{\textbf{Non-linear Scaling}}\\
During our superradiant decay the time evolution for all emitters is identical and thus correlated because all spins are coupled to the cavity mode equally. Therefore, the emitted radiation is coherent and interferes constructively leading to the typical $N^2​$ scaling of the emitted radiation intensity. This can be seen by rewriting the expression for the polarization in terms of the operators for the individual emitters
\begin{align}
\braket {S_+S_-} = \sum_i^N \braket{\sigma_+^i \sigma_-^i } + \sum_{i \ne j}^N\braket {\sigma_+^i\sigma_-^j}
\end{align}
The first sum gives the number of exictations and is proportional to the number of emitters ${N}$. The second sum accounts for interference terms and contains $N^2$ elements, responsible for the non-linear scaling of the emitted radiation intensity as given by $I\propto \frac{g^2}{\kappa}\braket{S_+S_-}$. Dephasing processes lead to differing phases for the $\sigma_+^i$ and $\sigma_-^j$ terms. For strong dephasing the second sum averages to zero and the intensity scales linearly with $N$ as in regular spontaneous emission with no superradiant burst observable. For an excited ensemble of emitters the coherence effects are negligible because the polarization term $\braket{\sigma_{+}^i\sigma_{-}^j}$ is zero. Half way through the decay process these terms become maximal, leading to the typical radiation intensity peak. Due to the fact that the experiment is carried out in a cavity with additional experimental imperfections such as misaligned magnetic fields for more than one subensemble in resonance with the cavity, this non-linear scaling exponent becomes smaller than two.

\newpage
\bibliographystyle{naturemag}
\bibliography{bibexport.bib}

\begin{thebibliography}{10}
\expandafter\ifx\csname url\endcsname\relax
  \def\url#1{\texttt{#1}}\fi
\expandafter\ifx\csname urlprefix\endcsname\relax\def\urlprefix{URL }\fi
\providecommand{\bibinfo}[2]{#2}
\providecommand{\eprint}[2][]{\url{#2}}

\bibitem{dicke_coherence_1954}
\bibinfo{author}{Dicke, R.~H.}
\newblock \bibinfo{title}{Coherence in {Spontaneous} {Radiation} {Processes}}.
\newblock \emph{\bibinfo{journal}{Physical Review}}
  \textbf{\bibinfo{volume}{93}}, \bibinfo{pages}{99--110}
  (\bibinfo{year}{1954}).

\bibitem{gross_superradiance:_1982}
\bibinfo{author}{Gross, M.} \& \bibinfo{author}{Haroche, S.}
\newblock \bibinfo{title}{Superradiance: {An} Essay on the Theory of Collective
  Spontaneous Emission}.
\newblock \emph{\bibinfo{journal}{Physics Reports}}
  \textbf{\bibinfo{volume}{93}}, \bibinfo{pages}{301--396}
  (\bibinfo{year}{1982}).

\bibitem{julsgaard_dynamical_2012}
\bibinfo{author}{Julsgaard, B.} \& \bibinfo{author}{M{\o}lmer, K.}
\newblock \bibinfo{title}{Dynamical Evolution of an Inverted Spin Ensemble in a
  Cavity: Inhomogeneous Broadening as a Stabilizing Mechanism}.
\newblock \emph{\bibinfo{journal}{Physical Review A}}
  \textbf{\bibinfo{volume}{86}}, \bibinfo{pages}{063810}
  (\bibinfo{year}{2012}).

\bibitem{mlynek_observation_2014}
\bibinfo{author}{Mlynek, J.~A.}, \bibinfo{author}{Abdumalikov, A.~A.},
  \bibinfo{author}{Eichler, C.} \& \bibinfo{author}{Wallraff, A.}
\newblock \bibinfo{title}{Observation of Dicke superradiance for two artificial
  atoms in a cavity with high decay rate}.
\newblock \emph{\bibinfo{journal}{Nature Communications}}
  \textbf{\bibinfo{volume}{5}} (\bibinfo{year}{2014}).

\bibitem{eschner_light_2001}
\bibinfo{author}{Eschner, J.}, \bibinfo{author}{Raab, C.},
  \bibinfo{author}{Schmidt-Kaler, F.} \& \bibinfo{author}{Blatt, R.}
\newblock \bibinfo{title}{Light interference from single atoms and their mirror
  images}.
\newblock \emph{\bibinfo{journal}{Nature}} \textbf{\bibinfo{volume}{413}},
  \bibinfo{pages}{495--498} (\bibinfo{year}{2001}).

\bibitem{devoe_observation_1996}
\bibinfo{author}{{DeVoe}, R.~G.} \& \bibinfo{author}{Brewer, R.~G.}
\newblock \bibinfo{title}{Observation of Superradiant and Subradiant
  Spontaneous Emission of Two Trapped Ions}.
\newblock \emph{\bibinfo{journal}{Physical Review Letters}}
  \textbf{\bibinfo{volume}{76}}, \bibinfo{pages}{2049--2052}
  (\bibinfo{year}{1996}).

\bibitem{scheibner_superradiance_2007}
\bibinfo{author}{Scheibner, M.} \emph{et~al.}
\newblock \bibinfo{title}{Superradiance of quantum dots}.
\newblock \emph{\bibinfo{journal}{Nature Physics}}
  \textbf{\bibinfo{volume}{3}}, \bibinfo{pages}{106--110}
  (\bibinfo{year}{2007}).

\bibitem{gross_observation_1976}
\bibinfo{author}{Gross, M.}, \bibinfo{author}{Fabre, C.},
  \bibinfo{author}{Pillet, P.} \& \bibinfo{author}{Haroche, S.}
\newblock \bibinfo{title}{Observation of Near-Infrared Dicke Superradiance on
  Cascading Transitions in Atomic Sodium}.
\newblock \emph{\bibinfo{journal}{Physical Review Letters}}
  \textbf{\bibinfo{volume}{36}}, \bibinfo{pages}{1035--1038}
  (\bibinfo{year}{1976}).

\bibitem{skribanowitz_observation_1973}
\bibinfo{author}{Skribanowitz, N.}, \bibinfo{author}{Herman, I.~P.},
  \bibinfo{author}{{MacGillivray}, J.~C.} \& \bibinfo{author}{Feld, M.~S.}
\newblock \bibinfo{title}{Observation of Dicke Superradiance in Optically
  Pumped {HF} Gas}.
\newblock \emph{\bibinfo{journal}{Physical Review Letters}}
  \textbf{\bibinfo{volume}{30}}, \bibinfo{pages}{309--312}
  (\bibinfo{year}{1973}).

\bibitem{inouye_superradiant_1999}
\bibinfo{author}{Inouye, S.} \emph{et~al.}
\newblock \bibinfo{title}{Superradiant {Rayleigh} {Scattering} from a
  {Bose}-{Einstein} {Condensate}}.
\newblock \emph{\bibinfo{journal}{Science}} \textbf{\bibinfo{volume}{285}},
  \bibinfo{pages}{571--574} (\bibinfo{year}{1999}).

\bibitem{ten_brinke_dicke_2015}
\bibinfo{author}{Ten~Brinke, N.} \& \bibinfo{author}{Sch\"utzhold, R.}
\newblock \bibinfo{title}{Dicke superradiance as a nondestructive probe for
  quantum quenches in optical lattices}.
\newblock \emph{\bibinfo{journal}{Physical Review A}}
  \textbf{\bibinfo{volume}{92}}, \bibinfo{pages}{013617}
  (\bibinfo{year}{2015}).

\bibitem{norcia_superradiance_2016}
\bibinfo{author}{Norcia, M.~A.}, \bibinfo{author}{Winchester, M.~N.},
  \bibinfo{author}{Cline, J. R.~K.} \& \bibinfo{author}{Thompson, J.~K.}
\newblock \bibinfo{title}{Superradiance on the millihertz linewidth strontium
  clock transition}.
\newblock \emph{\bibinfo{journal}{Science Advances}}
  \textbf{\bibinfo{volume}{2}} (\bibinfo{year}{2016}).

\bibitem{meiser_prospects_2009}
\bibinfo{author}{Meiser, D.}, \bibinfo{author}{Ye, J.},
  \bibinfo{author}{Carlson, D.~R.} \& \bibinfo{author}{Holland, M.~J.}
\newblock \bibinfo{title}{Prospects for a Millihertz-Linewidth Laser}.
\newblock \emph{\bibinfo{journal}{Physical Review Letters}}
  \textbf{\bibinfo{volume}{102}}, \bibinfo{pages}{163601}
  (\bibinfo{year}{2009}).

\bibitem{rose_coherent_2017}
\bibinfo{author}{Rose, B.} \emph{et~al.}
\newblock \bibinfo{title}{Coherent Rabi Dynamics of a Superradiant Spin
  Ensemble in a Microwave Cavity}.
\newblock \emph{\bibinfo{journal}{Physical Review X}}
  \textbf{\bibinfo{volume}{7}}, \bibinfo{pages}{031002} (\bibinfo{year}{2017}).

\bibitem{gross_maser_1979}
\bibinfo{author}{Gross, M.}, \bibinfo{author}{Goy, P.}, \bibinfo{author}{Fabre,
  C.}, \bibinfo{author}{Haroche, S.} \& \bibinfo{author}{Raimond, J.~M.}
\newblock \bibinfo{title}{Maser Oscillation and Microwave Superradiance in
  Small Systems of Rydberg Atoms}.
\newblock \emph{\bibinfo{journal}{Physical Review Letters}}
  \textbf{\bibinfo{volume}{43}}, \bibinfo{pages}{343--346}
  (\bibinfo{year}{1979}).

\bibitem{kaluzny_observation_1983}
\bibinfo{author}{Kaluzny, Y.}, \bibinfo{author}{Goy, P.},
  \bibinfo{author}{Gross, M.}, \bibinfo{author}{Raimond, J.~M.} \&
  \bibinfo{author}{Haroche, S.}
\newblock \bibinfo{title}{Observation of Self-Induced Rabi Oscillations in
  Two-Level Atoms Excited Inside a Resonant Cavity: The Ringing Regime of
  Superradiance}.
\newblock \emph{\bibinfo{journal}{Physical Review Letters}}
  \textbf{\bibinfo{volume}{51}}, \bibinfo{pages}{1175--1178}
  (\bibinfo{year}{1983}).

\bibitem{putz_protecting_2014}
\bibinfo{author}{Putz, S.} \emph{et~al.}
\newblock \bibinfo{title}{Protecting a spin ensemble against decoherence in the
  strong-coupling regime of cavity {QED}}.
\newblock \emph{\bibinfo{journal}{Nature Physics}}
  \textbf{\bibinfo{volume}{10}}, \bibinfo{pages}{720--724}
  (\bibinfo{year}{2014}).

\bibitem{temnov_superradiance_2005}
\bibinfo{author}{Temnov, V.~V.} \& \bibinfo{author}{Woggon, U.}
\newblock \bibinfo{title}{Superradiance and {Subradiance} in an
  {Inhomogeneously} {Broadened} {Ensemble} of {Two}-{Level} {Systems} {Coupled}
  to a {Low}-${Q}$ {Cavity}}.
\newblock \emph{\bibinfo{journal}{Physical Review Letters}}
  \textbf{\bibinfo{volume}{95}}, \bibinfo{pages}{243602}
  (\bibinfo{year}{2005}).

\bibitem{delanty_superradiance_2011}
\bibinfo{author}{Delanty, M.}, \bibinfo{author}{Rebić, S.} \&
  \bibinfo{author}{Twamley, J.}
\newblock \bibinfo{title}{Superradiance and phase multistability in circuit
  quantum electrodynamics}.
\newblock \emph{\bibinfo{journal}{New Journal of Physics}}
  \textbf{\bibinfo{volume}{13}}, \bibinfo{pages}{053032}
  (\bibinfo{year}{2011}).

\bibitem{jodoin_superradiance_1974}
\bibinfo{author}{Jodoin, R.} \& \bibinfo{author}{Mandel, L.}
\newblock \bibinfo{title}{Superradiance in an inhomogeneously broadened atomic
  system}.
\newblock \emph{\bibinfo{journal}{Physical Review A}}
  \textbf{\bibinfo{volume}{9}}, \bibinfo{pages}{873--884}
  (\bibinfo{year}{1974}).

\bibitem{bennett_phonon-induced_2013}
\bibinfo{author}{Bennett, S.~D.} \emph{et~al.}
\newblock \bibinfo{title}{Phonon-Induced Spin-Spin Interactions in Diamond
  Nanostructures: Application to Spin Squeezing}.
\newblock \emph{\bibinfo{journal}{Physical Review Letters}}
  \textbf{\bibinfo{volume}{110}}, \bibinfo{pages}{156402}
  (\bibinfo{year}{2013}).

\bibitem{lambert_superradiance_2016}
\bibinfo{author}{Lambert, N.} \emph{et~al.}
\newblock \bibinfo{title}{Superradiance with an ensemble of superconducting
  flux qubits}.
\newblock \emph{\bibinfo{journal}{Physical Review B}}
  \textbf{\bibinfo{volume}{94}} (\bibinfo{year}{2016}).

\bibitem{angerer_collective_2016}
\bibinfo{author}{Angerer, A.} \emph{et~al.}
\newblock \bibinfo{title}{Collective strong coupling with homogeneous {Rabi}
  frequencies using a {3D} lumped element microwave resonator}.
\newblock \emph{\bibinfo{journal}{Applied Physics Letters}}
  \textbf{\bibinfo{volume}{109}}, \bibinfo{pages}{033508}
  (\bibinfo{year}{2016}).

\bibitem{doherty_nitrogen-vacancy_2013}
\bibinfo{author}{Doherty, M.~W.} \emph{et~al.}
\newblock \bibinfo{title}{The nitrogen-vacancy colour centre in diamond}.
\newblock \emph{\bibinfo{journal}{Physics Reports}}
  \textbf{\bibinfo{volume}{528}}, \bibinfo{pages}{1--45}
  (\bibinfo{year}{2013}).

\bibitem{gruber_scanning_1997}
\bibinfo{author}{Gruber, A.} \emph{et~al.}
\newblock \bibinfo{title}{Scanning Confocal Optical Microscopy and Magnetic
  Resonance on Single Defect Centers}.
\newblock \emph{\bibinfo{journal}{Science}} \textbf{\bibinfo{volume}{276}},
  \bibinfo{pages}{2012--2014} (\bibinfo{year}{1997}).

\bibitem{bienfait_controlling_2016}
\bibinfo{author}{Bienfait, A.} \emph{et~al.}
\newblock \bibinfo{title}{Controlling spin relaxation with a cavity}.
\newblock \emph{\bibinfo{journal}{Nature}} \textbf{\bibinfo{volume}{531}},
  \bibinfo{pages}{74--77} (\bibinfo{year}{2016}).

\bibitem{astner_solid-state_2017}
\bibinfo{author}{Astner, T.} \emph{et~al.}
\newblock \bibinfo{title}{Solid-state electron spin lifetime limited by
  phononic vacuum modes}.
\newblock \emph{\bibinfo{journal}{arXiv:1706.09798 [cond-mat,
  physics:quant-ph]}}  (\bibinfo{year}{2017}).

\bibitem{tavis_exact_1968}
\bibinfo{author}{Tavis, M.} \& \bibinfo{author}{Cummings, F.~W.}
\newblock \bibinfo{title}{Exact Solution for an $N$-Molecule - Radiation-Field
  Hamiltonian}.
\newblock \emph{\bibinfo{journal}{Physical Review}}
  \textbf{\bibinfo{volume}{170}}, \bibinfo{pages}{379--384}
  (\bibinfo{year}{1968}).

\bibitem{e._m._purcell_proceedings_1946}
\bibinfo{author}{{E. M. Purcell}}.
\newblock \bibinfo{title}{Proceedings of the {American} {Physical} {Society}}.
\newblock \emph{\bibinfo{journal}{Physical Review}}
  \textbf{\bibinfo{volume}{69}}, \bibinfo{pages}{681} (\bibinfo{year}{1946}).

\bibitem{grezes_storage_2015}
\bibinfo{author}{Grezes, C.} \emph{et~al.}
\newblock \bibinfo{title}{Storage and retrieval of microwave fields at the
  single-photon level in a spin ensemble}.
\newblock \emph{\bibinfo{journal}{Physical Review A}}
  \textbf{\bibinfo{volume}{92}}, \bibinfo{pages}{020301}
  (\bibinfo{year}{2015}).

\bibitem{bohnet_steady-state_2012}
\bibinfo{author}{Bohnet, J.~G.} \emph{et~al.}
\newblock \bibinfo{title}{A steady-state superradiant laser with less than one
  intracavity photon}.
\newblock \emph{\bibinfo{journal}{Nature}} \textbf{\bibinfo{volume}{484}},
  \bibinfo{pages}{78--81} (\bibinfo{year}{2012}).

\bibitem{jin_proposal_2015}
\bibinfo{author}{Jin, L.} \emph{et~al.}
\newblock \bibinfo{title}{Proposal for a room-temperature diamond maser}.
\newblock \emph{\bibinfo{journal}{Nature Communications}}
  \textbf{\bibinfo{volume}{6}}, \bibinfo{pages}{8251} (\bibinfo{year}{2015}).

\bibitem{weiner_superradiant_2012}
\bibinfo{author}{Weiner, J.~M.}, \bibinfo{author}{Cox, K.~C.},
  \bibinfo{author}{Bohnet, J.~G.}, \bibinfo{author}{Chen, Z.} \&
  \bibinfo{author}{Thompson, J.~K.}
\newblock \bibinfo{title}{Superradiant Raman laser magnetometer}.
\newblock \emph{\bibinfo{journal}{Applied Physics Letters}}
  \textbf{\bibinfo{volume}{101}}, \bibinfo{pages}{261107}
  (\bibinfo{year}{2012}).

\bibitem{acosta_diamonds_2009}
\bibinfo{author}{Acosta, V.~M.} \emph{et~al.}
\newblock \bibinfo{title}{Diamonds with a high density of nitrogen-vacancy
  centers for magnetometry applications}.
\newblock \emph{\bibinfo{journal}{Physical Review B}}
  \textbf{\bibinfo{volume}{80}}, \bibinfo{pages}{115202}
  (\bibinfo{year}{2009}).

\bibitem{bienfait_reaching_2016}
\bibinfo{author}{Bienfait, A.} \emph{et~al.}
\newblock \bibinfo{title}{Reaching the quantum limit of sensitivity in electron
  spin resonance}.
\newblock \emph{\bibinfo{journal}{Nature Nanotechnology}}
  \textbf{\bibinfo{volume}{11}}, \bibinfo{pages}{253} (\bibinfo{year}{2016}).

\bibitem{amsuss_cavity_2011}
\bibinfo{author}{Ams{\"u}ss, R.} \emph{et~al.}
\newblock \bibinfo{title}{Cavity {QED} with Magnetically Coupled Collective
  Spin States}.
\newblock \emph{\bibinfo{journal}{Physical Review Letters}}
  \textbf{\bibinfo{volume}{107}}, \bibinfo{pages}{060502}
  (\bibinfo{year}{2011}).

\end{thebibliography}

\end{document}